\documentclass[9pt,twocolumn,twoside]{osajnl}

\journal{ol} 

\setboolean{shortarticle}{true}

\usepackage{amsmath}
\usepackage{amssymb}
\usepackage{amstext}
\usepackage{bm}
\usepackage{graphicx}
\usepackage{array}
\usepackage{multirow}
\usepackage{lipsum}
\usepackage{color}
\usepackage{ulem}

\title{Spatial mode detection by frequency upconversion}

\author[1]{Bereneice Sephton}
\author[1,*]{Adam Vall\'es}
\author[2,3]{Fabian Steinlechner}
\author[4,5]{Thomas~Konrad}
\author[6,7]{Juan~P.~Torres}
\author[1,8]{Filippus S. Roux}
\author[1]{Andrew Forbes}

\affil[1]{School of Physics, University of the Witwatersrand, Private Bag 3, Wits 2050, South Africa}
\affil[2]{Fraunhofer Institute for Applied Optics and Precision Engineering, Albert-Einstein-Str. 7, 07745 Jena, Germany}
\affil[3]{Friedrich Schiller University Jena, Abbe Center of Photonics, Albert-Einstein-Str. 6, 07745 Jena, Germany}
\affil[4]{School of Physics, University of KwaZulu-Natal, Durban, South Africa}
\affil[5]{National Institute of Theoretical Physics, UKZN Node, Durban, South Africa}
\affil[6]{ICFO-Institut de Ciencies Fotoniques, Barcelona Institute of Science and Technology, Mediterranean Technology Park, 08860 Castelldefels, Barcelona, Spain}
\affil[7]{Department of Signal Theory and Communications, Universitat Politecnica de Catalunya, Campus Nord D3, 08034 Barcelona, Spain}
\affil[8]{National Metrology Institute of South Africa, Meiring Naud\'e Road, Brummeria, Pretoria 0040, South Africa}

\affil[*]{Corresponding author: adam.vallesmari@wits.ac.za}

\doi{\url{https://doi.org/10.1364/OL.44.000586}}

\begin{abstract}
The efficient creation and detection of spatial modes of light has become topical of late, driven by the need to increase photon-bit-rates in classical and quantum communications. Such mode creation/detection is traditionally achieved with tools based on linear optics. Here we put forward a new spatial mode detection technique based on the nonlinear optical process of sum-frequency generation. We outline the concept theoretically and demonstrate it experimentally with intense laser beams carrying orbital angular momentum and Hermite-Gaussian modes. Finally, we show that the method can be used to transfer an image from the infrared band to the visible, which implies the efficient conversion of many spatial modes.
\end{abstract}

\setboolean{displaycopyright}{true}

\begin{document}

\maketitle

There has been tremendous development in methods to create and detect optical spatial modes, in particular complex structured light fields \cite{roadmap}, fueled by the desire for higher bit-rates in classical and quantum communication  \cite{Willner2015, wang2017data, Ndagano2018}. Employing a variety of concepts and implementations, the creation step may be achieved by refractive or diffractive field mapping, allowing lossless phase and amplitude modulation of an input beam~\cite{dickey2014laser}, or by single step modulation of the phase and/or amplitude. These approaches have been implemented by a variety of techniques: with dynamic phase by diffractive and free-form refractive optics, more commonly today with spatial light modulators (SLMs) \cite{White,SPIEbook}, by geometric phase using liquid crystal technology \cite{Marrucci2006}, or meta-materials and meta-surfaces \cite{Bomzon2002}. 

The detection step may be realized by running the aforementioned approaches in reverse, exploiting the reciprocity of light \cite{Forbes2016}. For example, if a particular hologram converts a Gaussian beam into some desired mode, then passing this mode through the same hologram in reverse produces a Gaussian beam, which may be coupled into a single mode fiber (SMF) to form a mode sensitive detector. This concept can be expanded to general phase-flattening \cite{mehul} and modal decomposition \cite{Flamm2012B,Flamm2013A} to detect any spatial state quantitatively.  Optical transformations may also be employed and have been successfully demonstrated for detection of Laguerre-Gaussian \cite{Berkhout2010A,LG}, Bessel-Gaussian \cite{dudley2013efficient} and Hermite-Gaussian (HG) modes \cite{HG}.
\begin{figure}
\centering \includegraphics[width=\linewidth]{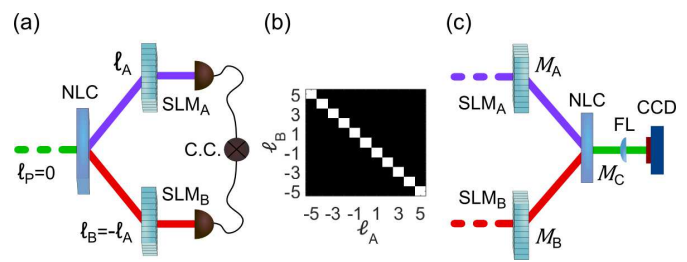} 
\caption{(a) In a traditional quantum experiment, a Gaussian mode pumps a nonlinear crystal (NLC) mediating the generation of two entangled photons with OAM values of $\ell$ and $-\ell$. (b)~Example of flat and anti-correlated modal spiral spectrum of the paired down-converted photons. (c) In the frequency up-conversion process, two incoming signals are engineered to be in specific states. The up-converted signal is detected in the far field, so that there is a non-zero signal only when the phases are conjugate.} 
\label{conceptfig}
\end{figure}

\begin{figure*}[t!]
\centering \includegraphics[width=\linewidth]{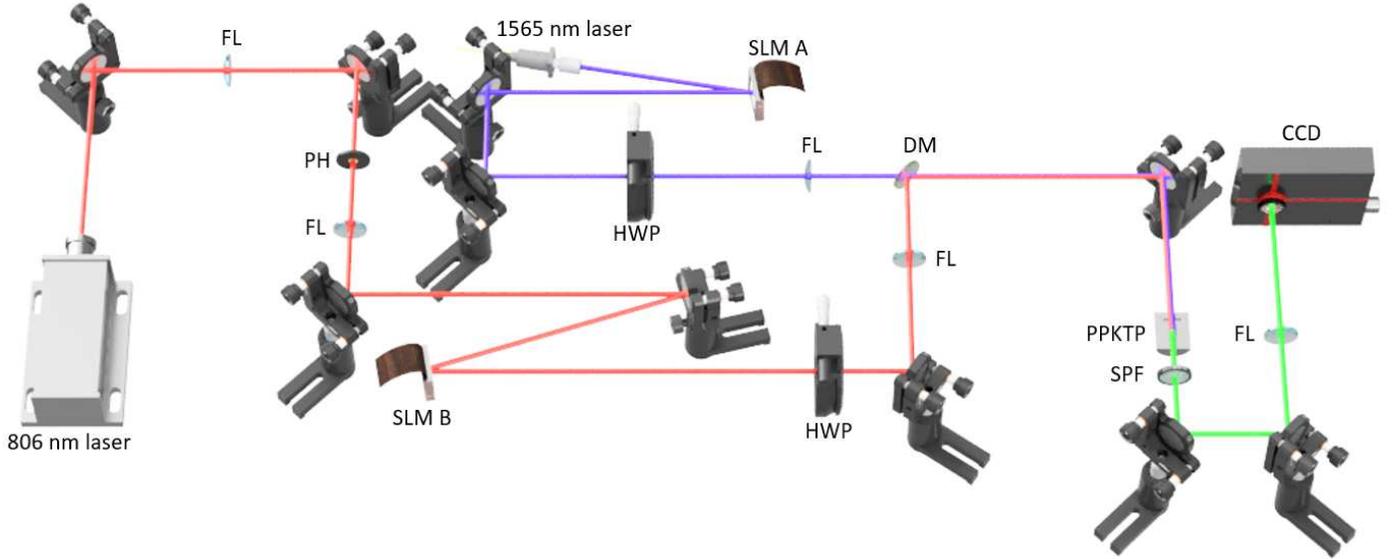} \caption{Schematic of the experimental up-conversion spatial mode detection setup. 806nm laser: high-power single frequency diode laser (Roithner); 1565 nm laser: single mode fiber pigtailed laser diode (Thorlabs); FL: Fourier lens; PH: pin hole; SLM: spatial light modulator (Holoeye); HWP: half-wave plate; DM: dichroic mirror; PPKTP: 5-mm-long nonlinear periodically-poled potassium titanyl phosphate crystal (Raicol); SPF: short-pass filter; CCD: CCD camera (Spiricon).} \label{4fimagingExpsetup}
\end{figure*}
 

While all the aforementioned approaches used linear optics for creation and detection, mode creation has also been demonstrated with nonlinear optics \cite{nl1, nl2, nl3, nl4, nl5, nl6, nl7, nl8, nl9}. With spontaneous parametric down-conversion (SPDC), the method of choice, one can create photon-pairs, entangled in their spatial degree of freedom~\cite{spdc}. It is instructive to outline how SPDC generates correlated modes, which we do here for a Gaussian pump and orbital angular momentum (OAM) modes \cite{OAM1} as an example. When a Gaussian beam pumps the SPDC process, it mediates the generation of paired entangled down-converted photons, signal ($A$) and idler ($B$), embedded into spatial modes satisfying $\ell_A = -\ell_B$, as depicted in Fig.~\ref{conceptfig}(a). The index $\ell_A$ and $\ell_B$  OAM modes characterized by an azimuthally varying phase of the form $\exp(i\ell \phi)$ with helicity $\ell$. Measuring the spatial state of photons $A$ and $B$ using the methods explained earlier, e.g. SLMs, one obtains the well-known anti-correlated spiral spectrum in OAM \cite{torres2003quantum}, as simulated in Fig.~\ref{conceptfig}(b). If SLM A is set to detect a $\ell = 2$ mode, then the detector in arm $B$ only ``clicks'' in coincidence with $A$ when SLM B is set to $\ell = -2$.

In this work we make use of the dual process, sum-frequency generation (SFG), for the detection and selection of spatial modes, as depicted in Fig.~\ref{conceptfig}(c). Signals $A$ and $B$ are up-converted to signal $C$, whose intensity in its central spot is detected in the far field. If $A$ and $B$ are OAM modes, a non-zero SFG signal is obtained only for $\ell_A + \ell_B = 0$.  The spatial modes of the light in $A$ and $B$ are controlled by SLMs, so the mode in $A$ can be determined precisely by a measurement on the up-converted signal in conjunction with a known spatial projection on~$B$. In this scheme the nonlinear crystal is the detector rather than the generator. We demonstrate the concept experimentally with both OAM and HG modes, as well as illustrate that the concept may be extended to arbitrary mode structures.  

We consider two input optical beams with spatial modes $M_A({\bf x})$ and $M_B({\bf x})$, that overlap at the nonlinear crystal to produce an up-converted beam with spatial mode $M_C({\bf x})$. In a paraxial thin-crystal approximation where we can neglect diffraction in the SFG process, the up-converted field can be calculated from the differential equation for SFG and can be expressed as
\begin{equation}
M_C({\bf x}) = g M_A({\bf x}) M_B({\bf x}) ,
\label{fsg0}
\end{equation}
where $g$ is a constant that represents the strength of the process. One can exploit the form of Eq.~(\ref{fsg0}) to implement a mode selector. To do so one installs a 2$f$-system behind the SFG crystal, which produces the Fourier transform of the output field in the back focal plane ${\cal M}_C({\bf u})={\cal F}\{M_C\}$. In the center of the back focal plane, one obtains
\begin{equation}
{\cal M}_C({\bf u}=0) \propto \int M_A({\bf x}) M_B({\bf x}) d^2 x ,
\label{fsg1}
\end{equation}
which has the form of an inner product. If one input beam is an unknown linear combination of orthogonal modes
\begin{equation}
M_A({\bf x}) = \sum_n C_n \Phi_n({\bf x}) ,
\label{fsg2}
\end{equation}
by preparing the other input beam to produce the complex conjugate of one particular mode in the crystal plane $M_B({\bf x}) = \Phi_m^*({\bf x})$, one can determine the modal content of the unknown beam
\begin{equation}
{\cal M}_C({\bf u}=0) \propto \sum_n C_n \int \Phi_n({\bf x}) \Phi_m^*({\bf x}) d^2 x = C_m.
\label{fsg3}
\end{equation}
The measured intensity in the center of the back focal plane would give $|C_m|^2$. This approach can be extended to implement a full state tomography of the unknown optical field \cite{tomo2}.

With the concept outlined, we now proceed to the experiment, shown schematically in Fig.~\ref{4fimagingExpsetup}. A 5-mm-long nonlinear periodically-poled potassium titanyl phosphate (PPKTP) crystal was pumped with two lasers ($A$ and $B$) at wavelengths of 1565 nm and 806 nm, respectively, so as to obtain up-converted light at a wavelength of 532 nm via SFG.  The crystal length was chosen to be 50$\times$ shorter than the Rayleigh length of the overlapping beams. We used a non-degenerate SFG type-0 process, which required vertically polarized inputs of different wavelengths, and gave us the highest nonlinear coefficient (quadratically proportional to the up-conversion efficiency). A 806 nm wavelength narrow band diode laser (Roithner) was used to generate one of the intense laser beams directed through the crystal, obtaining 50 mW of optical power after being spatially filtered with a 50-$\mu$m-diameter pinhole (PH) to form a collimated Gaussian mode. The spatially cleaned and horizontally polarized laser beam was then directed onto a SLM (Holoeye), allowing modulation of the spatial distribution and phase characteristics of the first diffracted order, creating our mode, $M_B$. After selection with an aperture, the first order was directed through the PPKTP crystal. A half-wave plate (HWP) changed the polarisation to vertical for the SFG process to occur. 

A pigtailed laser diode, tuned to a wavelength of $\lambda = 1565$~nm through temperature control, was collimated to produce a Gaussian mode of 50~mW optical power. The laser light was then modulated with a second SLM to create our mode $M_A$, the corresponding mode selected with an aperture and vertically polarized with a HWP. The two modulated laser beams were then combined with a dichroic mirror, allowing collinear SFG in the nonlinear crystal. A short-pass filter (SPF) separated the up-converted signal from the detectable pump wavelengths. The detected mode was either coupled into a SMF or, alternatively, the on-axis intensity on a CCD camera was measured.

\begin{figure}
	\centering \includegraphics[width=1\linewidth]{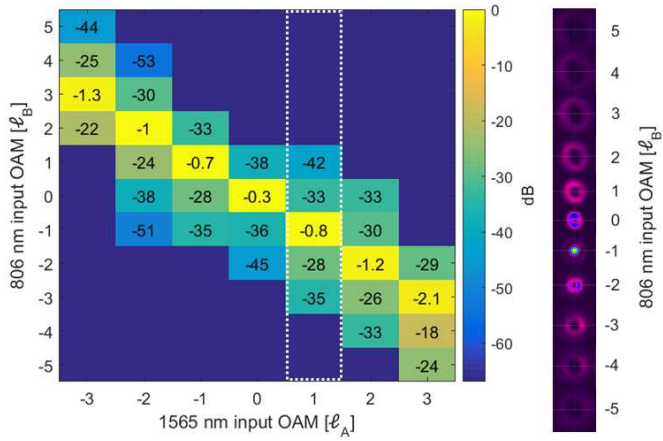} \caption{Cross-talk expressed in dB relative to the power in an input OAM mode with a particular helical charge for the 1565~nm ($\ell_A=[-3,3]$) and performing a modal decomposition by varying the helical charge of the input OAM mode for the 806~nm ($\ell_B=[-5,5]$). We have obviated the cross-talk values below -60 dB. The inset rightmost column shows the captured intensities for the OAM input mode $\ell_A = 1$, marked with a dotted white rectangle, to indicate where the central pixels are extracted.} \label{LGTomo}
\end{figure}

Two versions of the experimental setup were constructed: the first with the crystal in the Fourier plane of the SLMs and the second, to perform a high resolution up-conversion test, relaying the SLM plane into the crystal plane by using the proper imaging system. To work in the Fourier plane for the first setup configuration, a $f = 750$ mm lens was placed $f$-away from the SLM and the crystal. The resulting signal was measured in the far-field by placing a $f = 250$ mm lens after the crystal. Figure \ref{4fimagingExpsetup} depicts the first version of the setup used to obtain the mode detection results. In the second version, an $f = 100$ mm lens was then incorporated into a 4$f$-system before the crystal, relaying the SLMs plane into the crystal, allowing one to work with amplitude modulation generated images. A 4$f$-system was also constructed to image the crystal plane onto the CCD by adding a second $f = 300$ mm lens.

\begin{figure}
	\centering \includegraphics[width=1\linewidth]{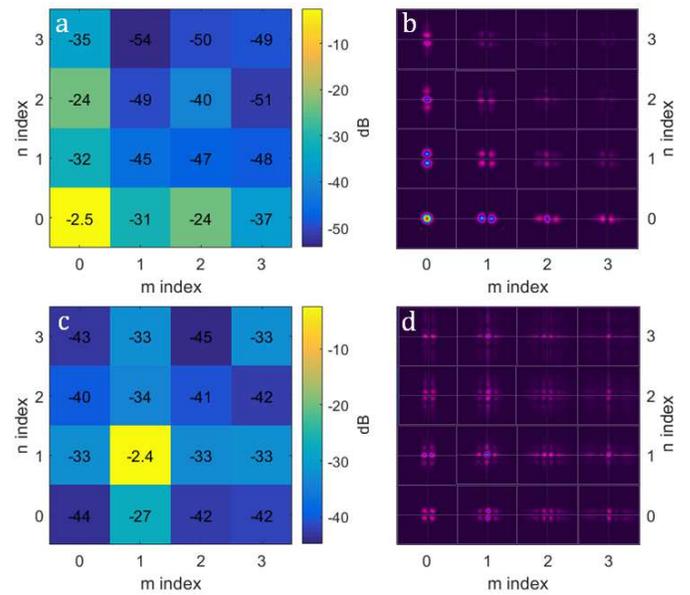} \caption{HG modal cross-talk given in dB for two different input HG$_{m,n}$ modes. The 1565~nm laser beam is encoded with (a) HG$_{0,0}$ and (c) HG$_{1,1}$, performing a modal decomposition with the 806~nm input laser beam such as HG$_{n,m}$, with the $m,n$ indexes ranging within [0,3]. We show the captures obtained in the right indicating the weightings of the detected modes in the left, by extracting the center pixel value in the intensity plots for (b) HG$_{0,0}$ and (d) HG$_{1,1}$. Dotted cross-hairs indicate the central coordinate.} 
    \label{HGdecomp}
\end{figure}

Our results shown in Figs.~\ref{LGTomo} and \ref{HGdecomp} confirm the concept of spatial mode detection by up-conversion using OAM and HG modes as examples. To quantify the quality of this spatial mode detection technique, a cross-talk matrix measurement was performed~\cite{li2014space}, showing how well we can distinguish between the desired input mode and the rest of the orthogonal modes. The measured cross-talk matrix, shown in Fig. \ref{LGTomo}, obeys the orthogonality relations with little cross-talk between neighboring spatial modes and validates the concept that a known projection in arm $B$ (on SLM B) together with a non-zero SFG ``signal'' results in a pattern sensitive detector for the mode in arm $A$. For instance, we see that when $M_A$ has OAM of $\ell_A = 1$, then a non-zero signal is found only when $\ell_B = -1$. Each column in Fig.~\ref{LGTomo} has been normalized with all its terms adding to unity, prior to expressing the cross-talk between the detected neighboring spatial modes in dB. The cross-talk measured values were found to be of -30 dB on average, having in the worst case scenario a value of -18 dB, being able also to easily isolate the correct input mode. The rightmost inset in Fig.~\ref{LGTomo} shows actual camera images, with a bright central spot only for the detected mode, with all other signals having practically no on-axis intensity. Note that we do not measure the radial properties of our OAM basis. Hence, we neglect the increase of radial modes when considering the up-conversion process of counter-rotating vortices \cite{n14}, as it plays no role in the OAM mode detection. While our OAM measurement was phase-only, phase and amplitude modes may be detected, which we demonstrate with HG modes in Fig.~\ref{HGdecomp}.

Figures~\ref{HGdecomp}(a) and (c) show the measurement outcome when a modal decomposition was performed on a 1565~nm laser beam with a particular encoded HG$_{m,n}$ input mode, with indexes (a)~$m=n=0$ and (c) $m=n=1$. The laser beam centered at 806~nm was used to perform a tomographic scan by varying the $m$ and $n$ indexes within the range [0,3], while measuring the resulting signal after the up-conversion process. Each cross-talk modal decomposition plot in Figs.~\ref{HGdecomp}(a) and (c) has been normalized with all its terms adding to unity, prior to being expressed in dB. A maximum on-axis intensity is obtained only when the 806~nm intense beam is encoded with the modes HG$_{0,0}$ and HG$_{1,1}$, as can be seen in Figs.~\ref{HGdecomp}(b) and (d), respectively. Here, since $M_A = \text{HG}_{0,0}$ (or $\text{HG}_{1,1}$), all on-axis intensities for $M_B$ are zero except for the case when $M_A = M_B^* = \text{HG}_{0,0}$ (or $\text{HG}_{1,1}$). The average efficiency of the process was $\approx 10^{-4}$, and further study is needed considering whether the efficiency is mode dependent.

\begin{figure}
	\centering \includegraphics[width=1\linewidth]{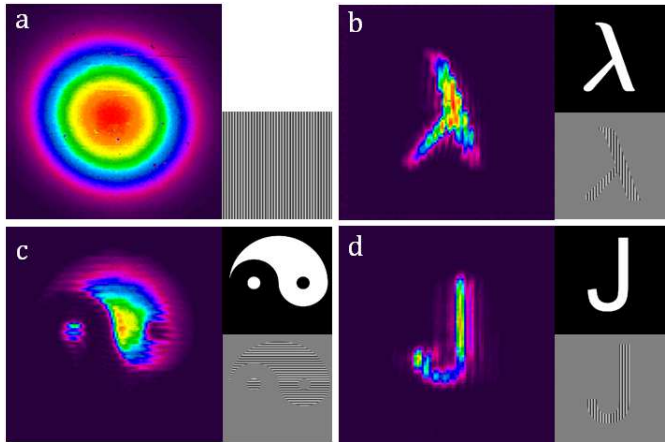} \caption{Up-converted images measured as $M_C$ with $M_A$ (1565~nm) as a Gaussian, shown in (a), and $M_B$ (806 nm) set to be (b) lambda, (c) yin yang and (d) jay symbols. The insets show the applied amplitude modulated mask and the corresponding hologram to imprint the image onto mode $M_B$. The emerging fringes are the Moir\'e pattern due to the finite resolution of the SLM screen.} \label{UpconFig}
\end{figure}

The extrapolation to more complex mode structures is limited only by the transverse resolution that can be obtained from the up-conversion process \cite{midwinter1968image, boyd1977infrared,vasilyev2012frequency}. In order to demonstrate the complexity of possible modes, we encoded various complex images in arm $B$, while keeping the other laser beam in arm $A$ as a Gaussian mode. Note that even if swapping the modes in $A$ and $B$, the resulting outcome would be the same. In Fig.~\ref{UpconFig} we show the resulting images after the up-conversion transfer, confirming the dynamic range of the technique, i.e., the technique works even with high resolution spatial modes, being able to transfer also the emerging Moir\'e pattern due to the finite resolution of the SLM screen. To test whether we were measuring only the up-converted 532 nm signal in all our results and not the residual 806 nm input (also detectable by the CCD camera), we changed the polarisation of the 1565 nm beam to horizontal, checking that the up-converted signal disappeared completely. This confirms a previously predicted result \cite{vasilyev2012frequency}.

In conclusion, we have outlined a simple concept for spatial mode detection by up-conversion and demonstrated it experimentally, using a variety of spatial mode structures. This experimental configuration paves the way for spatial mode detection of infrared band signals, but measuring them in the visible with cheaper and faster silicon based detector technology. The detection approach outlined here is based on mode projections and therefore not as efficient as mode sorters \cite{Berkhout2010A}, but nevertheless shows excellent modal discrimination. Even though we have demonstrated the technique with high intensity signals, the results are immediately applicable to single photon states too by employing single photon counting detectors \cite{dayan2005nonlinear}.

This work was supported by the Claude Leon Foundation. 


\end{document}